\documentclass[twocolumn, twocolappendix]{aastex7}

\usepackage{amsmath}

\shorttitle{}
\shortauthors{Zhu et al.}

\begin{document}

\title{Galaxy Underdensities Host the Clearest IGM Ly$\alpha$ Transmission and Indicate Anisotropic Reionization}

\author[0000-0003-3307-7525]{Yongda Zhu}
\thanks{JASPER Scholar}
\affiliation{Steward Observatory, University of Arizona, 933 North Cherry Avenue, Tucson, AZ 85721, USA}
\email[show]{yongdaz@arizona.edu}

\author[0000-0003-2344-263X]{George D. Becker}
\affiliation{Department of Physics \& Astronomy, University of California, Riverside, CA 92521, USA}
\email[]{georgeb@ucr.edu}

\author[0000-0003-2344-263X]{Anson D'Aloisio}
\affiliation{Department of Physics \& Astronomy, University of California, Riverside, CA 92521, USA}
\email[]{ansond@ucr.edu}

\author[0000-0003-4564-2771]{Ryan Endsley}
\affiliation{Department of Astronomy, The University of Texas at Austin, Austin, TX 78712, USA}
\email[]{ryan.endsley@austin.utexas.edu}

\author[0000-0002-1557-7422]{Nakul Gangolli}
\affiliation{Department of Physics \& Astronomy, University of California, Riverside, CA 92521, USA}
\email[]{ngang002@ucr.edu}

\author[0000-0001-9420-7384]{Christopher Cain}
\affiliation{School of Earth and Space Exploration, Arizona State University, Tempe, AZ 85287-6004, USA}
\email{clcain3@asu.edu}

\author[0000-0002-3407-1785]{Charlotte A. Mason}
\affiliation{Cosmic Dawn Center(DAWN), Denmark}
\affiliation{Niels Bohr Institute, University of Copenhagen, Jagtvej 128, 2200 Copenhagen N, Denmark}
\email{}

\author[0009-0008-7862-5277]{Seyedazim Hashemi}
\affiliation{Department of Physics \& Astronomy, University of California, Riverside, CA 92521, USA}
\email[]{}

\author[0009-0009-7741-7238]{Hui Hong}
\affiliation{Department of Physics \& Astronomy, University of California, Riverside, CA 92521, USA}
\email[]{}


\begin{abstract}
How galaxies drive reionization and what governs its geometry remain  fundamental questions. We present JWST/NIRCam wide-field slitless spectroscopy (WFSS) observations toward two of the most Ly$\alpha$-transmissive QSO sightlines near the end of reionization. We find that regions at $z \sim 5.7$ along both sightlines previously found to be low-density in Ly$\alpha$ emitters are also underdense in [O\,\textsc{iii}] emitters, with densities less than half the cosmic mean. Other transmissive regions, however, are found to coincide with average-density environments, indicating that multiple pathways may produce high IGM transmission. For the first time, we measure the two-dimensional cross-correlation between IGM transmission and galaxy positions, revealing evidence for anisotropic ionization geometry. Specifically, we detect enhanced transmission at transverse distances of $\Delta r \sim 0.8$ times the mean free path, consistent with ionizing photons escaping preferentially along large-scale structures that are aligned with, but offset from, the line of sight. This anisotropic escape may contribute to the observed patchiness of reionization and challenges the assumption of isotropic ionized bubble growth in current models.
\end{abstract}

\keywords{\uat{Reionisation}{1383}, \uat{High-redshift galaxies}{734}, \uat{Intergalactic medium}{813}}

\section{Introduction} \label{sec:intro}

One of the central open questions in cosmic reionization is how ionizing photons from the first galaxies escape into the intergalactic medium (IGM) and shape the evolving geometry of ionized regions. Observations of transmitted flux in the Ly$\alpha$ forest toward $z \sim 6$ QSOs \citep[e.g.,][]{fan_constraining_2006, mcgreer_model-independent_2015} suggest that reionization was largely complete by that epoch. However, the timing and geometry of reionization remain debated, with multiple lines of evidence supporting a late-ending and spatially inhomogeneous reionization process \citep[e.g.,][]{becker_evidence_2015, becker_evidence_2018, becker_damping_2024, bosman_new_2018, bosman_hydrogen_2022, mason_universe_2018, bolan_inferring_2022, yang_measurements_2020, zhu_chasing_2021, zhu_long_2022, zhu_damping_2024, spina_damping_2024, chen_impact_2025, umeda_probing_2025}.

Early evidence for such a late and patchy reionization scenario comes from the large sightline-to-sightline variations observed in the effective Ly$\alpha$ optical depth, $\tau_{\rm eff}$, at $z \sim 5$--6.\footnote{$\tau_{\rm eff} \equiv -\ln{\langle F \rangle}$, where $F$ is the continuum-normalized transmitted flux.} These fluctuations indicate significant variation in IGM transmission on scales of tens of comoving megaparsecs \citep[e.g.,][]{becker_evidence_2015, bosman_new_2018, yang_measurements_2020}. Such large-scale scatter may be driven by spatial variations in the ionizing ultraviolet background (UVB), the residual neutral hydrogen fraction, and/or thermal relics of reionization \citep{daloisio_large_2015, davies_large_2016,kulkarni_evolution_2019,nasir_observing_2020, keating_long_2020, garaldi_galaxy-igm_2024,garaldi_galaxy-igm_2025}. Different physical mechanisms predict different relationships between Ly$\alpha$ transmission and large-scale structure. For example, in models where fluctuations in $\tau_{\rm eff}$ are primarily driven by variations in a galaxy-driven UVB, transmissive regions are expected to correlate with galaxy overdensities \citep{davies_determining_2018,nasir_observing_2020}.  This scenario may be expected at $z \sim 5$--6 given the short mean free path of ionizing photons \citep{becker_evolution_2019, zhu_probing_2023}.

Observationally, several studies have found that highly opaque Ly$\alpha$ troughs are preferentially located in galaxy underdensities \citep[e.g.,][]{becker_evidence_2018, kashino_evidence_2020, christenson_constraints_2021, ishimoto_physical_2022}. This trend is generally consistent with fluctuating UVB models, including those wherein neutral islands persist below $z \sim 6$ \citep[e.g.,][]{kulkarni_evolution_2019, keating_long_2020, nasir_observing_2020}.  In contrast, the association between high opacity and low density is difficult to reconcile with models that attribute the scatter in Ly$\alpha$ transmission mainly to residual temperature fluctuations \citep[e.g.,][]{daloisio_large_2015}, which predict that recently reionized, low-density regions should appear more transmissive. Fluctuations in both UVB and temperature, as well as the presence of neutral islands, may all play a role in the relationship between density and opacity, however.

Surprisingly, \citet{christenson_relationship_2023} found that two of the most transmissive Ly$\alpha$ forest sightlines near the end of reionization ($z \sim 5.7$), toward the QSOs SDSS J1306+0356 (J1306) and PSOJ359--06 (J359), are located in galaxy underdensities traced by Ly$\alpha$ emitters (LAEs). This result challenges the expectations from UVB fluctuation models, which generally predict that low IGM Ly$\alpha$ opacity should correlate with galaxy overdensities, where the ionizing background is stronger. Instead, Christenson et al.\ showed that both high-opacity and low-opacity fields can lie in galaxy underdensities, although the association appears stronger at the high-opacity end. One possible explanation is that the ionizing background and gas temperature rise more rapidly in voids after reionization completes, leading to enhanced transmission. 
Follow-up modeling by \citet{gangolli_correlation_2025} using the FlexRT  simulations \citep{cain_flexrt_2024} found that such highly transmissive underdense sightlines can naturally occur if voids were reionized relatively late and remain hot, whereas transmissive regions in overdensities arise from locally enhanced ionizing backgrounds. 
Alternatively, LAEs may systematically avoid the highest-density peaks, as has been suggested in lower-redshift studies \citep[e.g.,][see also \citealp{scarlata_systematic_2025}]{cooke_nurturing_2013, huang_evaluating_2022}, possibly due to UVB suppression of star formation in dense environments \citep{bruns_clustering_2012, kashikawa_habitat_2007}.

While previous studies have primarily focused on one-dimensional correlations between Ly$\alpha$ opacity and galaxy density along sightlines \citep[e.g.,][]{kakiichi_jwst_2025}, such measurements inherently average over geometric information that may carry physical significance. In particular, they do not capture the relative spatial distribution of galaxies in the transverse and line-of-sight (LOS) directions, which could play a critical role if ionizing photons escape anisotropically or if ionized regions grow preferentially along large-scale structures. Measuring the IGM transmission as a function of both transverse and LOS separation from galaxies offers a more complete picture of how reionization proceeds spatially. 
In this way, the 2D cross-correlation builds directly on earlier opacity-density tests while extending them to probe the anisotropic geometry of reionization.

The observational studies noted above generally used wide-field narrow-band imaging to identify galaxies via their Ly$\alpha$ emission.  This approach has some advantages; galaxies are identified at a specific redshift, and the number density of galaxies around the QSO line of sight can be compared to the number density at the outskirts of the field, relieving some calibration uncertainties.  The visibility of Ly$\alpha$, however, will partly depend on the transparency of the foreground IGM, particularly if a foreground neutral island is present \citep[e.g.,][]{hashemi_ly_2025}.  Recently, a new approach to surveying galaxies along QSO lines of sight has been provided by JWST.  NIRCam \citep{rieke_performance_2023} wide-field slitless spectroscopy \citep[WFSS,][]{greene__2017} has been shown to efficiently identify galaxies at $z \simeq 5.3$--7.0 via their [O\,\textsc{iii}] emission \citep{kashino_eiger_2023,kashino_eiger_2025,jin_spectroscopic_2024,kakiichi_jwst_2025}.  This approach provides a complementary selection to Ly$\alpha$ emission while probing a longer redshift path, albeit over a narrow field of view than narrow-band imaging.

In this work, we present JWST/NIRCam WFSS observations toward the fields of \added{PSO~J359$-$06 (RA$=23^{\mathrm h}56^{\mathrm m}32.452^{\mathrm s}$, Dec$=-06^{\circ}22^{\prime}59.31^{\prime\prime}$; hereafter J359) and SDSS~J1306+0356 (RA$=13^{\mathrm h}06^{\mathrm m}08.259^{\mathrm s}$, Dec$=+03^{\circ}56^{\prime}26.19^{\prime\prime}$; hereafter J1306)}. Our first goal is test whether the transmissive regions at $z \sim 5.7$ are indeed low-density, as suggested by the Lyman-$\alpha$ emitter results \citep{christenson_relationship_2023}.  More broadly, we aim to use [O\,\textsc{iii}] emitters (O3Es) to investigate the conditions that produce highly transmissive regions near the end of reionization. To probe the spatial relationship between galaxies and IGM opacity in greater detail, we move beyond traditional one-dimensional statistics and introduce a new two-dimensional stacking approach that quantifies the IGM transmission as a function of both transverse and LOS separation from galaxies. This technique enables us to investigate the geometry of ionized structures and to test whether anisotropic ionizing photon escape contributes to shaping the Ly$\alpha$ forest transmission. By comparing our results to mock sightlines from cosmological simulations, we assess whether current reionization models reproduce the observed geometry of transmission near galaxies.

This paper is organized as follows. In Section~\ref{sec:data}, we describe the QSO spectra and JWST observations. Section~\ref{sec:results} presents the O3Es identified in both fields. We discuss the implications for reionization geometry in Section~\ref{sec:discussion}, and summarize in Section~\ref{sec:summary}. We adopt a flat $\Lambda$CDM cosmology with $\Omega_{\rm m}=0.3$, $\Omega_\Lambda=0.7$, and $h=0.7$. Comoving distances are used unless otherwise stated.

\begin{figure*}[!ht]
    \centering
    \includegraphics[width=0.49\linewidth]{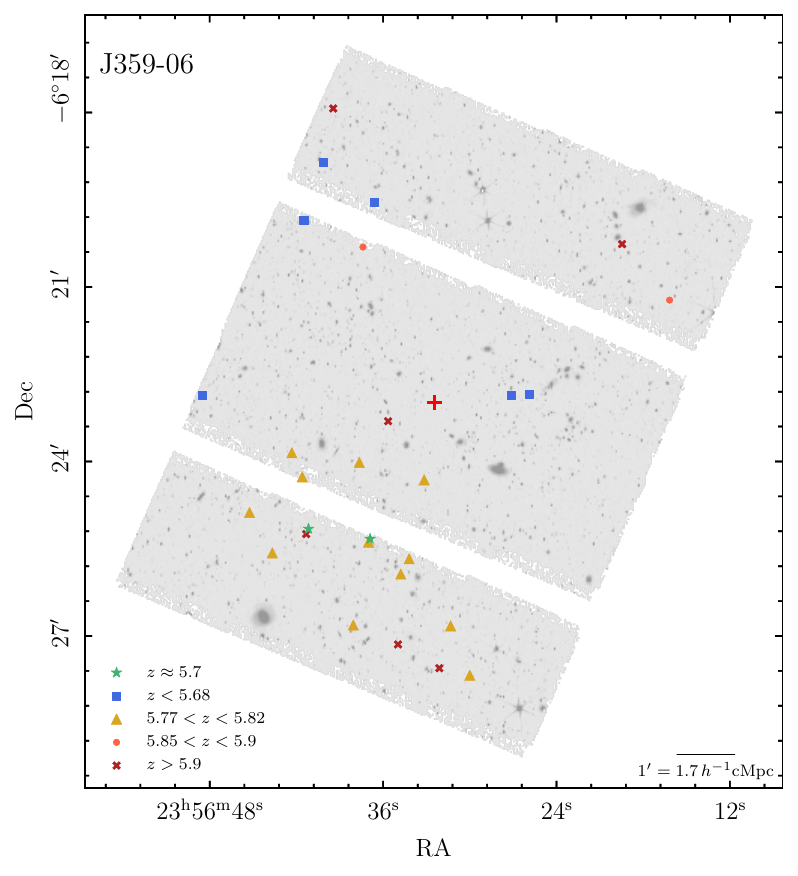}
    \includegraphics[width=0.49\linewidth]{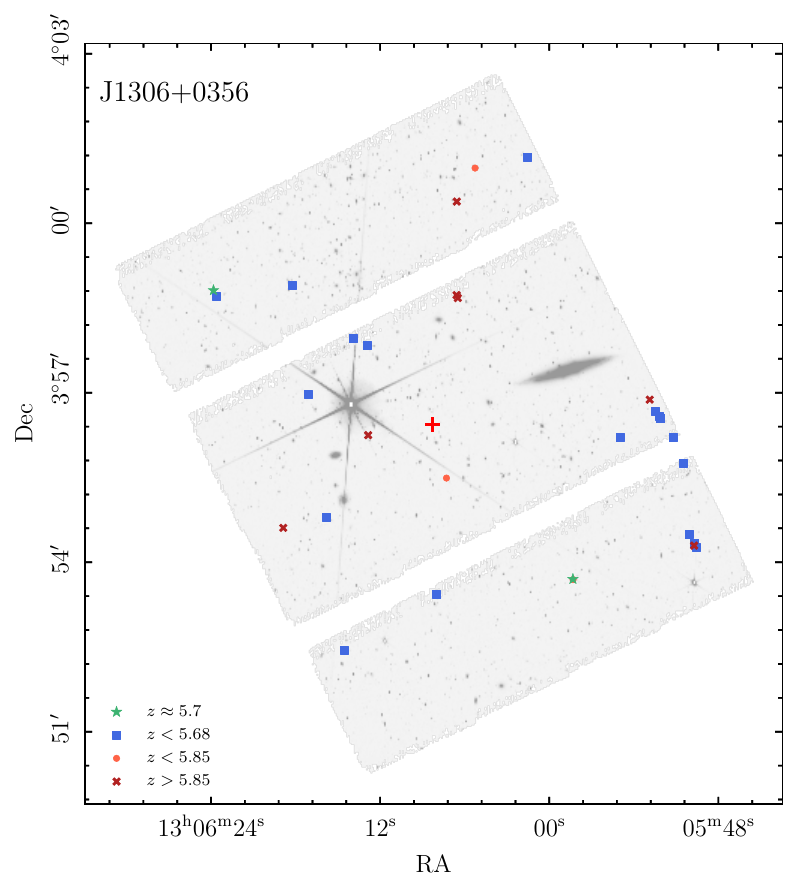}
    \caption{
    Sky distribution of [O\,\textsc{iii}] emitters overlaid on JWST/NIRCam F335M mosaics for the J359$-$06 (left) and J1306$+$0356 (right) fields. Symbols indicate spectroscopic redshift bins: green stars mark galaxies at $z \approx 5.7$, corresponding to the most transmissive Ly$\alpha$ forest region as identified by the HSC/NB816 filter in \citet{christenson_relationship_2023}; blue squares show galaxies at $z < 5.68$; yellow triangles denote the $5.77 < z < 5.82$ overdensity in J359$-$06, visible in the southeast quadrant and associated with strong metal absorption (C\,\textsc{iv}, Si\,\textsc{iv}) in the QSO spectrum; red circles and dark red crosses represent galaxies at $z > 5.85$, with crosses indicating sources excluded from the galaxy-IGM cross-correlation analysis (see text for details). The background QSO in each field is marked with a red plus. Notably, the regions near the most transmissive sightlines (green stars) exhibit a deficit of [O\,\textsc{iii}] emitters.}
    \label{fig:image}
\end{figure*}

\begin{figure*}[!ht]
    \centering
    \includegraphics[width=1.0\linewidth]{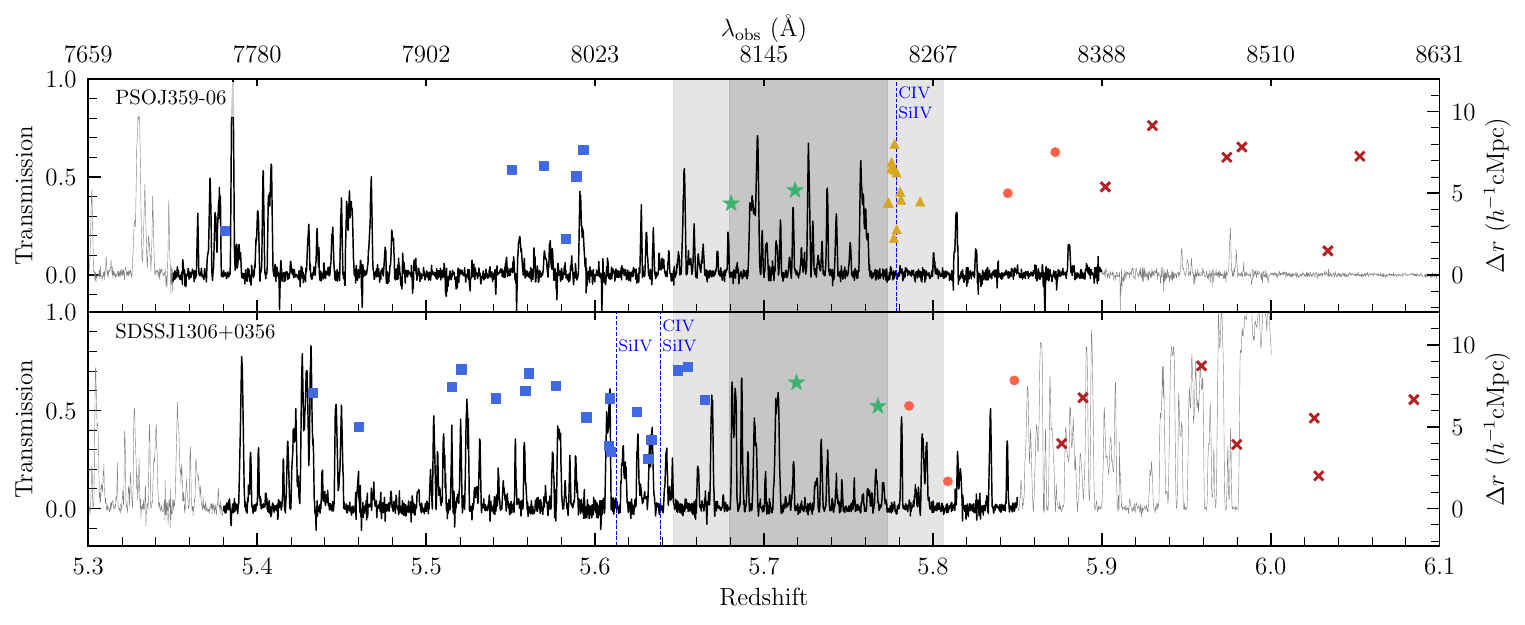}
    \caption{
Ly$\alpha$ forest transmission spectra of J359$-$06 (top) and J1306$+$0356 (bottom), shown as a function of redshift (bottom axis) and observed wavelength (top axis). The black curves indicate normalized flux, with the thicker portions marking the redshift ranges used for the IGM-galaxy cross-correlation analysis. Symbols mark the redshifts of [O\,\textsc{iii}] emitters and their transverse distance to the sightline, using the same color and shape coding as in Figure~\ref{fig:image}. Green stars highlight sources near $z \approx 5.7$. The dark gray shaded regions correspond to the NB816 filter coverage as used in \citet{christenson_relationship_2023}, coinciding with the most transmissive segment in each sightline. The lighter gray bands show $\pm 25\ h^{-1}$~cMpc regions centered on the NB816 peak redshift. Blue dashed lines indicate the locations of metal absorbers (C\,\textsc{iv}, Si\,\textsc{iv}) identified in the XQR-30 spectra.}
    \label{fig:spec}
\end{figure*}

\section{Data}\label{sec:data}
\subsection{QSO spectra}
The two QSO sightlines examined in this work, J1306 and J359, represent two of the most Ly$\alpha$-transparent lines of sight currently known at $z \sim 5.7$ \citep{christenson_relationship_2023}. J359 has a systemic redshift of $z = 6.1718$ derived from [C~\textsc{ii}] emission \citep{eilers_detecting_2020}, and was observed as part of the XQR-30 Large Program with VLT/X-Shooter \citep{dodorico_xqr-30_2023}. J1306 has a redshift of $z = 6.0330$ based on the [C~\textsc{ii}] 158~$\mu$m line \citep{venemans_kiloparsec-scale_2020}, and was also observed with X-Shooter in archival programs \citep{becker_evolution_2019}. Both spectra are drawn from the larger sample of $z > 5.5$ QSOs compiled by \citet{zhu_chasing_2021,zhu_long_2022}, which includes a homogeneous reduction of data from Keck/ESI and VLT/X-Shooter.

\citet{christenson_relationship_2023} characterized the IGM transmission along these sightlines using the effective Ly$\alpha$ optical depth. To enable direct comparison with Ly$\alpha$ emitter (LAE) studies, they evaluated $\tau_{\rm eff}$ over $50\ h^{-1}$~cMpc windows centered at 8177~\AA\ ($z \simeq 5.7$), and reported $\tau_{\rm eff}^{50} = 2.617 \pm 0.009$ for J1306 and $\tau_{\rm eff}^{50} = 2.661 \pm 0.009$ for J359. These measurements place both sightlines among the most transmissive known at this redshift (cf. the global $\tau_{\rm eff} \sim 3.665$ at $z=5.7$ in \citealp{bosman_hydrogen_2022}). Also, these sightlines are free from long dark gaps in the forest \citep{zhu_chasing_2021,zhu_long_2022}.

To match the redshift range probed by narrowband galaxy selection with the HSC/NB816 filter, \citet{christenson_relationship_2023} also computed $\tau_{\rm eff}$ over a $28\ h^{-1}$~cMpc window corresponding to the FWHM of the narrow band filter. They found $\tau_{\rm eff}^{28} = 2.475 \pm 0.010$ for J1306 and $\tau_{\rm eff}^{28} = 2.392 \pm 0.009$ for J359. These values confirm that the high transmission persists over the scales relevant for mapping galaxy environments in our NIRCam observations.

\subsection{NIRCam Observations}

We conducted JWST/NIRCam WFSS observations of the J1306 and J359 fields as part of Program 4092 (PI: G.~Becker, \citealp{becker_how_2023}), 
\footnote{\added{All the JWST data used in this paper can be found in MAST under \dataset[doi:10.17909/4z7r-ef97]{\doi{10.17909/4z7r-ef97}}.}}
focusing on [\ion{O}{3}]$\lambda\lambda4960,5008$ emission from galaxies at $z = 5.68$--$5.77$, corresponding to the peak Ly$\alpha$ forest transmissivity. We selected the F335M medium-band filter for grism spectroscopy to balance spectral resolution and background suppression, enhancing sensitivity to emission lines while mitigating spectral overlap. Simultaneous short-wavelength imaging in F115W provides source localization and rest-frame UV measurements.

Each field was observed with a $4 \times 2$ NIRCam mosaic, covering approximately $10 \times 7.5$~arcmin$^2$. We employed the INTRAMODULEBOX4 dither pattern to fill detector gaps, ensure uniform coverage, and facilitate artifact rejection. We note that, however, the WFSS data excludes a narrow strip centered on the QSO in each field (see the completeness map in Appendix \ref{sec:completeness}). Observations were conducted in the GRISM R configuration. Exposure parameters were optimized using the JWST Exposure Time Calculator (ETC) to reach the line sensitivity required to detect \added{typical} star-forming galaxies at $z \sim 5.7$ \added{(corresponding to $\log_{10}(L_{5008}/{\rm erg\,s^{-1}})\simeq 42.1$, or an effective $5\sigma$ line-flux-density limit of $f_\lambda \simeq 1.5\times10^{-19}\,{\rm erg\,s^{-1}\,cm^{-2}\,\AA^{-1}}$ at $\lambda_{\rm obs}\approx 3.36\,\mu{\rm m}$)}. For the WFSS exposures (F115W + F335M), we used SHALLOW4 readout mode with 10 groups per integration. Direct imaging (F115W + F335M) were then taken with 5 groups per integration. Each exposure consists of a single integration, yielding a total exposure time of $\sim$2100~s per grism pointing.

\subsection{NIRCam Data Reduction}

For the direct imaging data, we follow the methodology described in \citet{endsley_jwstnircam_2023}, using the JWST Science Calibration Pipeline (v1.11.3) with custom modifications. Snowball and wisp artifacts were removed by applying sky flats and wisp templates derived from publicly available data. The 1/f noise and two-dimensional background subtraction were performed on an amplifier-by-amplifier basis using the {\tt sep} package \citep{barbary_sep_2016}, applied to the *\_cal.fits files.

The NIRCam/WFSS data were reduced using the publicly available grism reduction pipeline from \citet{sun_first_2023}, with updated trace and calibration files. We briefly summarize the key steps below.

Each exposure was first processed with the standard JWST Stage 1 calibration pipeline (v1.11.2), which performs bias subtraction, dark current correction, and ramp fitting to produce count-rate images. Pixel-level flat-field corrections and two-dimensional sky background subtraction were then applied. World Coordinate System (WCS) solutions were assigned and refined by cross-matching bright stars in the short-wavelength direct images with Gaia DR3 catalog sources \citep{gaia_collaboration_gaia_2023}, achieving an astrometric precision better than 0\farcs05.

Spectral extraction was performed using the wavelength solution, trace models, and sensitivity curves from \citet{sun_first_2023}. For each grism exposure, we computed the two-dimensional (2D) dispersed spectrum of every cataloged object based on its location in the direct image, accounting for geometric distortion and spectral tilt. These 2D spectra were interpolated onto a common wavelength-spatial grid and coadded to improve signal-to-noise.

Finally, we extracted one-dimensional (1D) spectra using a fixed-width boxcar aperture of 5 pixels ($\sim$0\farcs3) along the spatial direction. Local background and continuum levels were estimated and subtracted using median filtering. These 1D and 2D spectra form the basis for our emission-line measurements and galaxy redshift identification.

\section{Results}\label{sec:results}

\begin{deluxetable*}{ccccccc}
\tablenum{1}
\tabletypesize{\footnotesize}
\tablecaption{[O\,\textsc{iii}] Emitters Around Highly Transmissive QSO Sightlines}
\tablehead{
\colhead{Field} &
\colhead{ID} &
\colhead{RA} &
\colhead{DEC} &
\colhead{$z$} &
\colhead{$\Delta r$ [cMpc/$h$]} &
\colhead{$L_{5008}$ [$10^{42}$ erg s$^{-1}$]}
}
\decimalcolnumbers
\startdata
J359 & 752 & 359.12499 & -6.46125 & 5.78 & 8.01 & 7.1 $\pm$ 0.2 \\
J359 & 863 & 359.13381 & -6.45923 & 5.98 & 7.83 & 8.7 $\pm$ 0.2 \\
J359 & 1323 & 359.14571 & -6.45243 & 5.97 & 7.20 & 5.4 $\pm$ 0.2 \\
J359 & 1716 & 359.13054 & -6.44710 & 5.78 & 6.52 & 5.2 $\pm$ 0.1 \\
J1306 & 1146 & 196.56055 & 3.87408 & 5.56 & 7.18 & 3.0 $\pm$ 0.2 \\
J1306 & 2291 & 196.53332 & 3.89048 & 5.46 & 5.01 & 3.6 $\pm$ 0.2 \\
J1306 & 2762 & 196.49296 & 3.89488 & 5.79 & 6.28 & 2.3 $\pm$ 0.2 \\
J1306 & 2778 & 196.49301 & 3.89504 & 5.77 & 6.25 & 1.8 $\pm$ 0.2 \\
\enddata
\tablecomments{
Columns:
(1) Background QSO field;
(2) \added{Galaxy ID};
(3) \& (4) Coordinates in J2000;
(5) Spectroscopic redshift;
(6) Projected transverse distance from the QSO sightline in comoving Mpc/$h$;
(7) Observed [O\,\textsc{iii}] $\lambda5008$ line luminosity with $1\sigma$ uncertainty, in units of $10^{42}$ erg s$^{-1}$.\\
(The complete table is available in machine-readable format.)
}
\label{tab:o3emitters}
\end{deluxetable*}

\subsection{[O\,\textsc{iii}] Emitters}

We identify [O\,\textsc{iii}] emitter candidates in the extracted 1D spectra following the methodology of \citet{wang_spectroscopic_2023}. Specifically, we search for [O\,\textsc{iii}]~$\lambda5008$ emission lines with signal-to-noise ratio (S/N)~$\geq 5$, and require the line width to lie within $0.5 \times \mathrm{FWHM}_\mathrm{res} \leq \mathrm{FWHM} \leq 7 \times \mathrm{FWHM}_\mathrm{res}$, where the nominal spectral resolution for the F335M grism is $\mathrm{FWHM}_\mathrm{res} \approx 180~\mathrm{km~s^{-1}}$. We additionally require detection of the corresponding [O\,\textsc{iii}]~$\lambda4960$ line at $S/N > 2$, and reject candidates with unphysical line ratios (with ratios $\frac{\mathrm{S/N}_{5008}}{\mathrm{S/N}_{4960}}$ outside 1.5 to 4).

All selected candidates were visually inspected by two team members (YZ and GDB). During this step, we also compared the morphology of each source in the 2D emission-line maps with its appearance in the direct imaging, in order to remove blended sources and potential interlopers. Only high-confidence emitters confirmed independently by both inspectors were included in the final sample.

This selection yields 28 O3Es in the J359 field and 30 in the J1306 field. Table~\ref{tab:o3emitters} lists the positions, redshifts, [O\,\textsc{iii}] luminosities, and projected transverse distances from the QSO sightlines. The spatial distribution of the emitters is shown in Figure~\ref{fig:image}

Our selection reaches a 40\% completeness limit of $\log_{10}(L_{5008}/[\mathrm{erg~s^{-1}}]) \approx 42.1$, and becomes 80\% complete by $\log_{10}(L_{5008}) \approx 42.3$. The overall contamination rate is low, typically below 10\%. Details of the completeness and contamination estimates are provided in Appendix~\ref{sec:completeness}.

\subsection{Galaxy Density}

\begin{figure}[!ht]
    \centering
    \includegraphics[width=\linewidth]{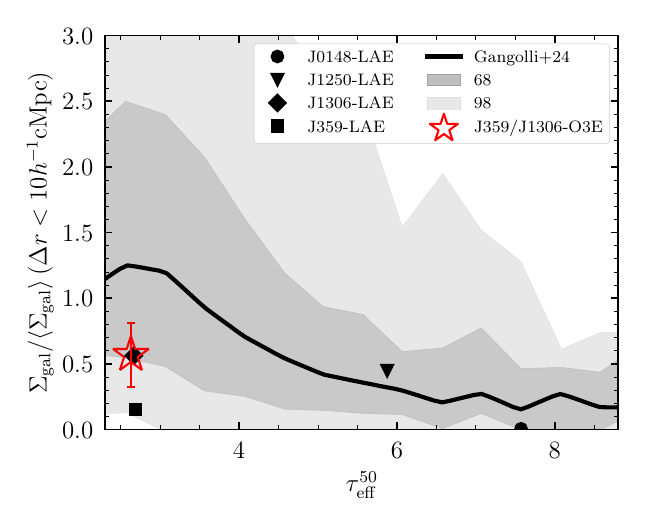}
    \caption{
    Relation between Ly$\alpha$ forest opacity and galaxy density at $z \sim 5.7$. 
    The red star marks our measurement from the J359 and J1306 fields based on [O\,\textsc{iii}] emitters (\added{shown as a single point combining two fields}). 
    Black symbols show LAE-based measurements from \citet{christenson_constraints_2021}. 
    The black curve and shaded bands denote the mean relation and scatter from the ``$\dot{N}_{\rm ion} \propto L_{\rm UV}$'' \textsc{FlexRT} model of \citet{gangolli_correlation_2025}.
    }
    \label{fig:tau_density}
\end{figure}

Figure~\ref{fig:spec} shows distribution of O3Es alongside the Ly$\alpha$ forest transmission. We detect 28 emitters in J359 and 30 in J1306. However, within the $28~\mathrm{cMpc}/h$ redshift range corresponding to the NB816 filter, which traces the most transmissive region of the Ly$\alpha$ forest, only 2 O3Es are found in each field.

We compare our observed O3E counts to expectations from published luminosity functions. Each NIRCam field probes a comoving volume of approximately $1.2 \times 10^5~\mathrm{cMpc}^3$ over $z = 5.3$ to 6.1 \footnote{We only consider regions with completeness greater than 0.}. Convolving the \citet{matthee_eiger_2023} [O\,\textsc{iii}] luminosity function (O3LF) with our median completeness curve and luminosity threshold ($\log L_{5008} > 42.1$) yields a predicted count of $106.4^{+38.7}_{-38.7}$ emitters per field. In comparison, the FRESCO-based O3LF at $z \sim 7.1$ from \citet{meyer_jwst_2024} predicts $34.0^{+56.6}_{-34.0}$. Our observed values, 28 and 30 emitters in J359 and J1306 respectively, are consistent with the Meyer et al. prediction but fall below the Matthee et al. estimate, which is based on the J0100 field known to host a luminous QSO and giant galaxy overdensities.

Focusing on the most transmissive IGM regions, defined by the NB816 redshift interval spanning approximately $1.4 \times 10^4~\mathrm{cMpc}^3$, we detect only two emitters per field. This is lower than the expected number of $\sim 3.5$ from the field average, \added{ with an uncertainty of $\pm 2.1$ when combining Poisson noise and a 0.30 fractional cosmic variance estimated from mock sightlines (1$\sigma$; see Section~\ref{sec:simulation}). Considering both fields together, we detect four emitters in total, compared to an expectation of $\sim 7.0 \pm 3.0$ under the same assumptions, placing the observed value at the lower bound of the 1$\sigma$ range.} \added{Therefore, for the combined J359/J1306 sample we infer $\Sigma_{\rm gal}/\langle\Sigma_{\rm gal}\rangle = 0.57 \pm 0.245$ (1$\sigma$), which we show in the Ly$\alpha$ opacity--galaxy density relation in Figure~\ref{fig:tau_density}.}
This is also lower than the $12.4^{+4.5}_{-4.5}$ and $4.0^{+6.6}_{-4.0}$ emitters expected from the \citet{matthee_eiger_2023} and \citet{meyer_jwst_2024} luminosity functions, respectively, under the same luminosity threshold \added{after correcting for our survey completeness}. The resulting emitter density in the peak transmission zones is therefore approximately half of the field-wide mean, and even more strongly suppressed compared to either luminosity function, including the higher-redshift $z \sim 7$ case. This underdensity further supports the interpretation that these exceptionally transmissive sightlines trace galaxy-sparse regions.

We also identify a striking overdensity at $z = 5.78$, composed of 12 O3Es. These galaxies are coincident in the redshift space with high-ionization metal absorbers, including Si\,\textsc{iv} and C\,\textsc{iv}, detected in the QSO absorption spectra from the XQR-30 survey \citep[see Appendix~\ref{sec:metal};][]{davies_xqr-30_2023}. The column densities $\log N_\mathrm{Si\,IV} = 12.68$ and $\log N_\mathrm{C\,IV} = 13.26$ suggest highly ionized and enriched gas, potentially in the circum-galactic medium of an intervening galaxy, likely maintained by local sources of ionizing radiation.  Moreover, the $z = 5.78$ overdensity lies approximately $23~\mathrm{cMpc}/h$ along the line of sight from the blue end of the NB816 transmission window ($z \sim 5.7$), exceeding the ionizing photon mean free path at this redshift ($\lambda_{\mathrm{mfp}} \sim 12~\mathrm{cMpc}/h$). It is therefore unlikely to be the dominant cause of the enhanced Ly$\alpha$ transmission.

Our O3E measurements confirm the results of \citet{christenson_relationship_2023}, who find that these highly transmissive sightlines trace underdensities of LAEs. \added{We note, however, that our WFSS configuration has reduced (and near-zero) completeness within a small region centered on the QSO position for $z\simeq5.6$--5.8, due to gaps in grism spectral coverage (see Appendix~\ref{sec:completeness}, Figure~\ref{fig:comp2d}).} \added{Our result} is illustrated in Figure~\ref{fig:tau_density}, which places our O3E-based opacity-density measurement alongside LAE results from \citet{christenson_constraints_2021} and predictions from the \textsc{FlexRT} simulations \citep[see Section \ref{sec:simulation} for details;][]{gangolli_correlation_2025}. In such low-density regions, the lower gas densities may naturally produce lower opacities, provided the local UV background is not strongly suppressed.  If these regions have also been recently reionized then the elevated gas temperatures may lower the recombination rate, further reducing the opacity \citep{daloisio_large_2015}.  At face value, our results are in some tension with those of \citet{kashino_eiger_2025}.  They also used O3Es to trace the relationship between IGM transmission and galaxy density, finding a broadly positive trend at $z \sim 5.7$ \citep[see also][]{jin_spectroscopic_2024,kakiichi_jwst_2025}.  While this is consistent with the underdensities in LAEs observed near opaque sightlines \citep[e.g.,][]{becker_evidence_2018,kashino_evidence_2020,christenson_constraints_2021}, it runs contrary to our finding that at least some of the most transmissive sightlines have low galaxy densities.  \added{Because the completeness gap is confined to a small area 
\footnote{\added{The low-completeness gap spans $\simeq 7.5\times1.0~\mathrm{arcmin}^2$ centered on the QSO, corresponding to a comoving volume of $\simeq 6\times10^2~(h^{-1}\mathrm{cMpc})^3$ over the NB816 line-of-sight path-length of $\simeq 28~h^{-1}\mathrm{cMpc}$.}}
around the QSO position, it primarily affects density estimates at the smallest projected separations, and we therefore interpret the O3E density near the line of sight as a lower limit.} \added{Importantly, our mock observations adopt the same NIRCam footprint (including the central completeness gap), so the comparison between the observed and expected galaxy counts remains valid within the quoted uncertainties.} One possible explanation is that previous studies have probed only sightlines with average to high opacity, and do not include the lowest-opacity regions such as those surveyed here.

At the same time, we note that the transmissive region at $z=5.5$-5.65 toward J1306 coincides with an environment of at least average galaxy density. This highlights that transmissive sightlines do not arise exclusively in underdense regions, consistent with the large scatter at low $\tau_{\rm eff}$ expected in models by \citet{gangolli_correlation_2025}. Such diversity suggests that multiple physical pathways may lead to high IGM transmission, and motivates a more detailed spatial analysis. \added{Importantly, our conclusion that the underdensity probed is consistent with the independent LAE constraints from Subaru/HSC \citep{christenson_relationship_2023}, suggesting that any galaxies missed in the WFSS completeness gap are unlikely to dominate the inferred large-scale environment.} We also caution that the opacity-density relation can evolve with redshift.  In the mock data we explore in Appendix \ref{sec:convergence}, the transmission-galaxy signal strengthens rapidly toward higher redshift.  Part of the difference between the average-density transmissive regions we observe at $z \sim 5.5$--5.65 and the low-density transmissive regions at $z \sim 5.7$ may therefore reflect redshift evolution \citep[see also][]{kashino_eiger_2025}.

\section{COMPARISON TO MODELS}\label{sec:discussion}

Our main finding, that at least some of the most transmissive sightlines coincide with galaxy underdensities, has previously been compared to reionization models \citep[e.g.,][]{christenson_relationship_2023}. Here we extend these efforts with a more detailed statistical analysis, using forward-modeled mock observations from the FlexRT simulations to examine both one- and two-dimensional correlations.

\subsection{Mock Observations from Simulations} \label{sec:simulation}

We construct mock observations using the FlexRT simulations \citep{cain_flexrt_2024} presented by \citet{gangolli_correlation_2025}, which perform post-processed radiative transfer on a $200~h^{-1}~\mathrm{cMpc}$ cosmological volume. The simulation uses $2\times 2048^3$ dark matter and gas particles and resolves halos down to $M_\mathrm{min} = 3 \times 10^9~h^{-1}~M_\odot$ at $z = 5.7$. Radiative transfer is carried out on a uniform $200^3$ grid to model spatial variations in temperature and ionization state.
Halo catalogs are linked to galaxy UV luminosities ($M_{\rm UV}$) through abundance matching.
In these simulations, reionization ends late at $z \sim 5.0$-5.3, with a global neutral fraction of $x_{\mathrm{HI}} \sim 20\%$ at $z = 6$. The emissivity is calibrated to reproduce the Ly$\alpha$ forest mean transmission measured by \citet{bosman_hydrogen_2022}, and the resulting evolution of the ionizing photon mean free path is broadly consistent with \citet{becker_mean_2021}, \citet{zhu_probing_2023}, and \citet{gaikwad_measuring_2023}.

Following \citet{zhu_chasing_2021}, we generate mock Ly$\alpha$ forest spectra along 50 more transmissive sightlines out of 100 simulated sightlines and extract galaxy catalogs within apertures matching our NIRCam observations. We do not attempt to model [O\,\textsc{iii}] emitters directly; instead, we select mock galaxies using a rest-UV magnitude threshold, with the minimum halo mass adjusted so that the number of mock galaxies matches the observed [O\,\textsc{iii}] emitter sample. \added{For reference, the minimum halo mass we selected at $z=5.6$ is $\sim 1.9\times10^{11}\,M_\odot$, corresponding to a rest-UV selection threshold of $M_{\rm UV}<-20.7$ in the simulation.}
To account for stochastic sampling of galaxies along the sightlines, we create multiple realizations by randomly drawing galaxies from these catalogs when constructing mock observations.
Tests with halo occupation fractions below unity show results that are effectively degenerate with raising the halo mass threshold and therefore do not affect our conclusions. Appendix~\ref{sec:convergence} explores the effects of redshift evolution, field of view, and depth variations.

\added{In the FlexRT models, under-densities around highly transmissive sightlines at $z\simeq5.7$ become more common when the volume-averaged neutral fraction is lower. For the J359-like criterion of observing $\le 1$ LAE within $R=10\,{\rm cMpc}/h$, the fraction of mock sightlines is 3\% in the $x_{\mathrm{HI}}=15\%$ model and 9\% in the $x_{\mathrm{HI}}=5\%$ model \citep{gangolli_correlation_2025}.}
This suggests that the lowest-opacity end of the opacity-density relation may serve as a sensitive probe of the final stages of reionization and the thermal state of cosmic voids. 
While the specific models explored by \citet{christenson_relationship_2023} could not reproduce both transmissive and opaque sightlines simultaneously, the FlexRT models of \citet{gangolli_correlation_2025} perform substantially better, capturing the observed scatter in the opacity-density relation and providing a more consistent framework for interpreting both regimes.
Our comparison suggests that reconciling the full range of observations will likely require models that combine several effects, including spatial fluctuations in ionizing emissivity, reionization timing, and residual temperature structure.

\subsection{One-Dimensional Opacity-Density Relation}

To characterize the spatial relationship between galaxies and Ly$\alpha$ forest transmission along QSO sightlines, we compute the 1D IGM--galaxy relation. As shown in Figure~\ref{fig:spec}, we select spectral windows that exclude the proximity zone (in J1306) and match the redshift range of low-redshift [O\,\textsc{iii}] emitters (blue squares). Specifically, we evaluate the normalized transmission contrast, defined as $T/\langle T \rangle - 1$, as a function of distance to the nearest galaxy, where distance is expressed in units of the ionizing photon mean free path ($\lambda_{\rm mfp}$; \citealp{becker_mean_2021,zhu_probing_2023}) at the redshift of each O3E. This statistic reflects whether IGM transmission is locally enhanced or suppressed near galaxies.  Similar statistics have been used in prior studies as a diagnostic of reionization geometry and timing \citep[e.g.,][]{kakiichi_jwst_2025,kashino_eiger_2025}.

\begin{figure}[!ht]
    \centering
    \includegraphics[width=1.0\linewidth]{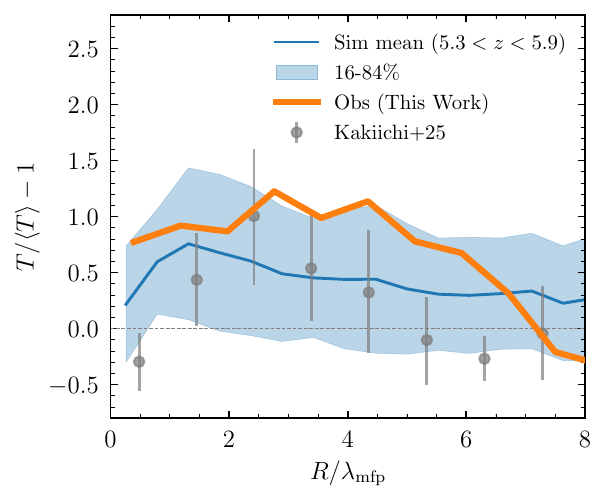}
    \caption{
    Stacked one-dimensional cross-correlation between normalized transmission and distance to the nearest galaxy, shown as $T/\langle T \rangle - 1$ versus $R / \lambda_{\mathrm{mfp}}$. The blue curve shows the mean prediction from the \textsc{FlexRT} simulation over $5.3 < z < 5.9$ for the intermediate-luminosity and limited-FOV mock galaxy selection. The shaded band indicates the 16--84th percentile range across mock realizations. The orange curve shows our observed measurement combining the J1306+0356 and J359--06 sightlines. Gray points with error bars are reproduced from \citet{kakiichi_jwst_2025}. 
    }
    \label{fig:1d_opacity_density}
\end{figure}

Figure~\ref{fig:1d_opacity_density} shows the measured 1D relation from the combined J359 and J1306 sightlines (orange line), compared to results from \citet{kakiichi_jwst_2025} (gray points), who analyzed [O\,\textsc{iii}] emitters from the ASPIRE program. For theoretical context, we include the mean and 16th-84th percentile range from 10,000 mock realizations of two sightlines drawn from the FlexRT simulation described in Section~\ref{sec:simulation}.
These mock realizations are selected from the 50 most transmissive simulated sightlines, which typically have $\tau_{\mathrm{eff}} < 3$, comparable to the observed values. 
We verified that restricting to even more transmissive sightlines ($\tau_{\mathrm{eff}} < 2.7$) yields consistent results but reduces the sampling volume of the simulation.
Because the number of observed [O\,\textsc{iii}] emitters is small, we do not attempt to bootstrap observational uncertainties directly; instead, the shaded region reflects the expected variance estimated from the simulation ensemble.

Our observed relation broadly agrees with both the simulation and the results from ASPIRE; the transmission is slightly lower within approximately one mean free path of galaxies, and peaks at separations of about 2--6 mean free paths \added{(taking $\lambda_{\rm mfp}\simeq 17\,h^{-1}\,{\rm cMpc}$ at $z\simeq5.6$ in \citealp{zhu_probing_2023}, this corresponds to $\sim 17\,h^{-1}\,{\rm cMpc}$ and $\sim 34$--$102\,h^{-1}\,{\rm cMpc}$, respectively)}. This trend suggests that the immediate surroundings of galaxies may still contain dense gas that suppresses Ly$\alpha$ transmission, while regions at moderate separations likely benefit from escaping ionizing radiation that maintains high IGM transparency.

This result is qualitatively consistent with \citet{garaldi_galaxy-igm_2024}, who found that galaxy-IGM cross-correlations for transmissive sightlines peak at larger distances than those for opaque sightlines. In their interpretation, this offset reflects more extended and mature ionized regions due to a longer mean free path at the end stages of reionization. Although we do not divide our sample by transmission class, the peak observed at 2-6 mean free paths in our high-transmission sightlines may reflect a similar underlying structure.

Despite these insights, the 1D IGM-galaxy relation has important limitations. The small number of available sightlines (only two in our case) makes the statistic sensitive to sample variance and shot noise. Moreover, the relation depends on a degenerate mix of gas density, photoionization rate $\Gamma_{\mathrm{HI}}$, and thermal state, which cannot be disentangled without additional information. As shown in \citet{gangolli_correlation_2025}, even models with very different reionization histories can yield similar 1D correlations once calibrated to match the observed mean flux.
At first glance, this behavior may seem at odds with the opacity-density relation (Figure~\ref{fig:tau_density}), which shows that low opacities  coincide with galaxy underdensities. In practice, however, the two statistics probe different regimes: the opacity-density relation isolates the environments of individual sightlines, whereas $T(R)$ averages over all pixels at fixed separation, which at large $R$ primarily sample typical-density IGM rather than voids. The two approaches are therefore complementary rather than contradictory, with each highlighting different aspects of the galaxy-IGM connection.


\begin{figure*}
    \centering
    \includegraphics[width=0.8\linewidth]{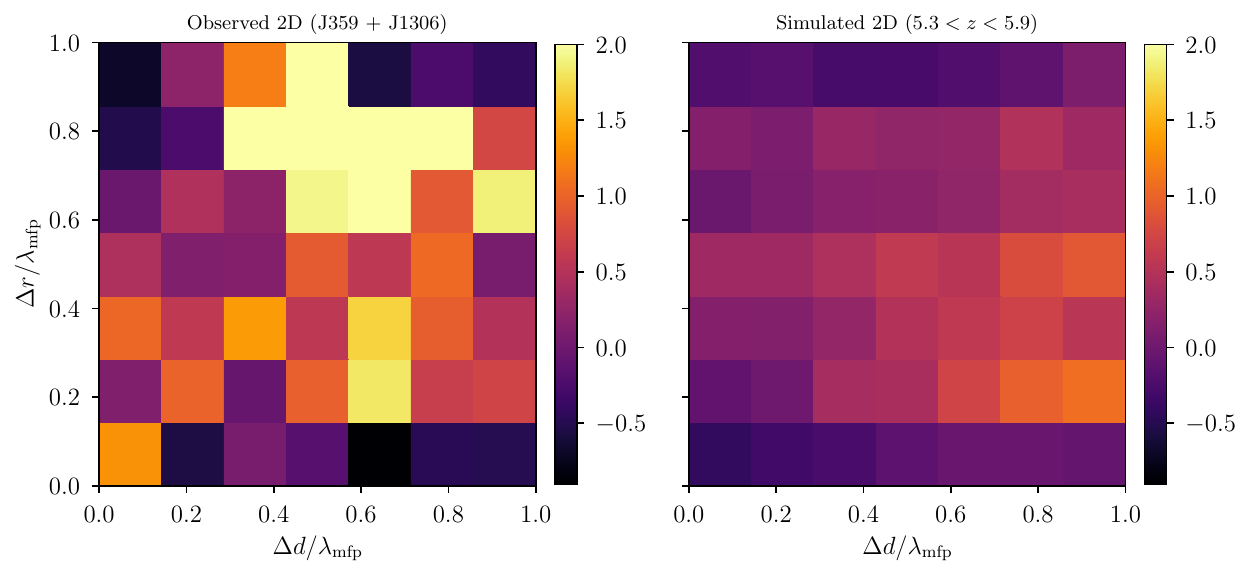}
    \caption{
    Two-dimensional transmission maps. \textbf{Left:} Observed map of normalized Ly$\alpha$ transmission, shown as $(T / \langle T \rangle) - 1$, as a function of transverse separation $\Delta r$ and line-of-sight separation $\Delta d$ from galaxies, both normalized by the mean free path ($\lambda_{\rm mfp}$). A prominent horizontal band of enhanced transmission is visible at $\Delta r \sim 0.8\,\lambda_{\rm mfp}$, suggesting anisotropic structure in the IGM. \textbf{Right:} Corresponding map from mock sightlines generated from the \textsc{FlexRT} simulation, constructed by stacking 10,000 bootstrap realizations in the redshift range $5.3 < z < 5.9$. The simulated enhancement appears more isotropic, with the strongest signal near the lower-right quadrant.}
    \label{fig:2d_obs}
\end{figure*}

\begin{figure*}
    \centering
    \includegraphics[width=0.8\linewidth]{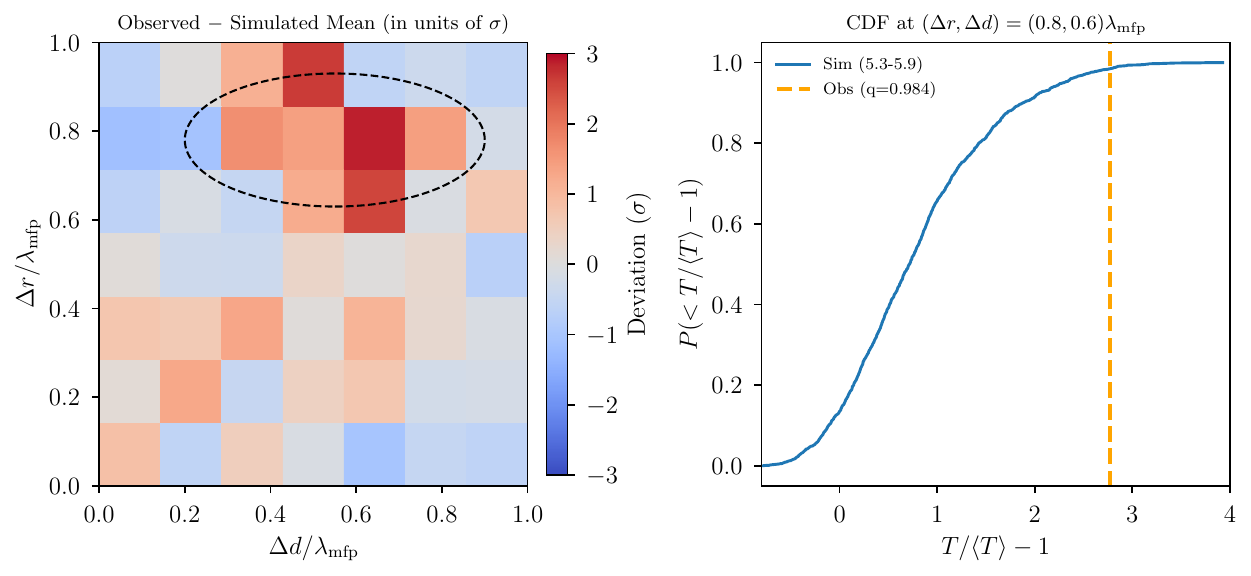}
    \caption{\textbf{Left:} Significance map showing the difference between the observed 2D transmission map and the mean of the mock ensemble, in units of standard deviation ($\sigma$) based on 10{,}000 bootstrap realizations. A prominent off-axis enhancement is detected at $(\Delta r, \Delta d) \sim (0.8, 0.6)\, \lambda_{\mathrm{mfp}}$, reaching $\sim 2.5\sigma$ significance (highlighted by the dashed ellipse), indicating a significant excess of IGM transmission compared to the simulations. 
\textbf{Right:} Cumulative distribution function (CDF) of simulated transmission values at the bin centered on $(\Delta r, \Delta d) = (0.8, 0.6)\, \lambda_{\mathrm{mfp}}$. The vertical orange dashed line marks the observed value, which lies at the 98.4th percentile ($q = 0.984$), confirming the anomalous nature of the observed enhancement.
}
    \label{fig:sigma_map}
\end{figure*}

\subsection{Two-Dimensional Transmission Structure}

To further examine the spatial relationship between galaxies and Ly$\alpha$ transmission, we construct a 2D map of the normalized Ly$\alpha$ transmission, $T/\langle T \rangle - 1$, as a function of transverse separation $\Delta r$ and LOS separation $\Delta d$ from each galaxy. Both axes are expressed in units of $\lambda_{\mathrm{mfp}}$ at the galaxy redshift, interpolated from measurements in \citet{zhu_probing_2023}. We stack all pixel-galaxy pairs across the J359 and J1306 fields, including only separations within $1\,\lambda_{\mathrm{mfp}}$ in both directions to minimize edge effects and ensure statistical robustness.

The observed 2D transmission map (Figure~\ref{fig:2d_obs}, left panel) reveals a prominent horizontal band of enhanced transmission at $\Delta r \sim 0.8\,\lambda_{\mathrm{mfp}}$, offset from the sightline. This feature suggests that ionizing radiation may preferentially escape along directions not aligned with the line of sight, indicative of anisotropic structure in the surrounding IGM.

To assess the statistical significance of this feature, we generate 10,000 mock 2D maps by repeatedly resampling galaxies along the 50 transmissive simulated sightlines from the \textsc{FlexRT} simulation (see Section~\ref{sec:simulation}). The simulated mean map (Figure~\ref{fig:2d_obs}, right panel) shows a much more isotropic enhancement pattern, with the enhanced transmission located near the lower-right quadrant, at low $\Delta r$ and moderate $\Delta d$. This contrast suggests that the observed enhancement may not be easily reproduced in standard reionization models.

We quantify the discrepancy by computing a significance map of the residuals between the observed and simulated mean maps (Figure~\ref{fig:sigma_map}, left panel). The enhancement near $(\Delta r, \Delta d) \sim (0.8, 0.6)\,\lambda_{\mathrm{mfp}}$ exceeds $2.5\sigma$ significance, marked by the dashed ellipse. A cumulative distribution function (CDF) analysis of mock values at this grid point confirms the anomaly, placing the observed value at the 98.4th percentile (Figure~\ref{fig:sigma_map}, right panel).
In the mock ensemble, the probability that a single map cell exceeds $2.5\sigma$ is $\simeq 3.3\%$.

\subsection{Potential Evidence of Anisotropic Reionization Geometry}

The simulated transmission enhancement at small transverse separation ($\Delta r \lesssim 0.5 \lambda_{\rm mfp}$) and moderate LOS separation ($\Delta d \sim 0.4$--0.6\,$\lambda_{\mathrm{mfp}}$) persists even after removing the field-of-view limitation (see Appendix \ref{sec:convergence}). This indicates that the feature is not a geometric artifact, but instead reflects a physical effect: the clustering of galaxies along the LOS. In this scenario, multiple galaxies aligned in redshift contribute overlapping ionized regions, enhancing Ly$\alpha$ transmission without requiring anisotropic emission from individual sources. This phenomenon, known as \textit{LOS clustering}, has been proposed elsewhere. For instance, \citet{chen_jwst_2024} suggest that galaxy groups extended along the LOS may enhance Ly$\alpha$ visibility by producing elongated ionized tunnels in the IGM (see also \citealp{hashemi_ly_2025,chen_impact_2025}).

However, LOS clustering alone cannot explain the observed off-axis enhancement at large transverse separation ($r \sim 0.8\,\lambda_{\mathrm{mfp}}$) and moderate LOS distance ($\Delta d \sim 0.6\,\lambda_{\mathrm{mfp}}$). This structure suggests that ionized regions around galaxies may not be spherically symmetric. Instead of isotropic bubbles, the geometry appears elongated or filamentary, potentially shaped by large-scale structure and direction-dependent ionizing photon transmission.

\begin{figure*}[!ht]
    \centering
    \includegraphics[width=0.95\textwidth]{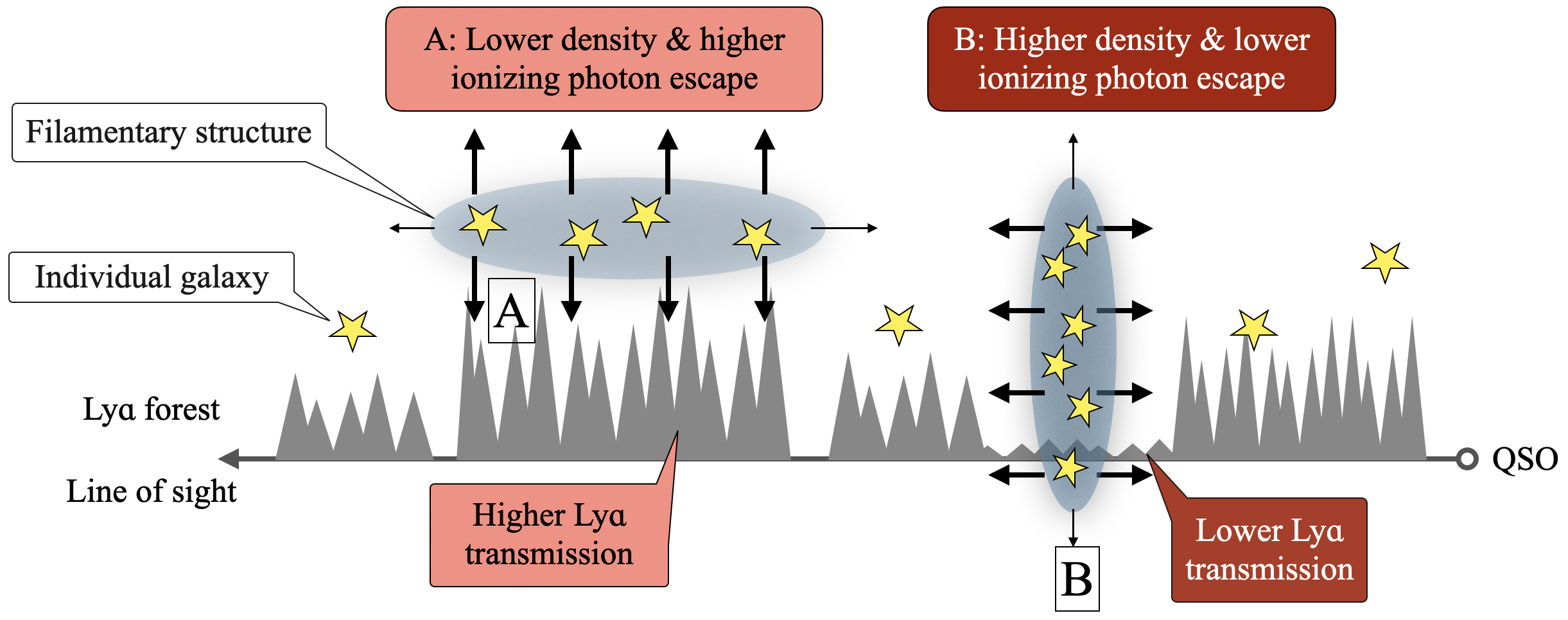} 
    \caption{
    Illustration of anisotropic ionizing photon escape and its effect on Ly$\alpha$ transmission. The schematic shows two potentially representative environments along a quasar line of sight (horizontal axis). In region \textbf{A}, ionizing photons preferentially escape vertically from a lower-density, filamentary structure offset from the sightline, producing an elongated ionized channel that overlaps the Ly$\alpha$ forest and yields \emph{higher} transmission. In region \textbf{B}, a denser structure intersects the sightline directly; ionizing photons escape primarily along the transverse direction, resulting in a \emph{lower} Ly$\alpha$ transmission in the forest. Together, these cases illustrate how directional photon escape and local geometry can generate anisotropic ionized structures even in similar large-scale environments.
    }
    \label{fig:cartoon}
\end{figure*}

We hypothesize that enhanced transmission may preferentially occur when galaxies cluster in structures aligned with but offset from the QSO sightline, allowing ionized channels created by galaxies to overlap with the Ly$\alpha$ forest (see Figure \ref{fig:cartoon} for a schematic illustration). In contrast, clustering along transverse paths may mean that the QSO sightline intersects denser, self-shielded material, reducing transmission. This geometry is supported by the spectra shown in Figure~\ref{fig:spec}, where galaxies roughly aligned with the sightline often coincide with transmission spikes, while those intersect the sightline do not. 

This interpretation supports a view of reionization shaped by directional emissivity, anisotropic escape paths, and cosmic web geometry. Rather than isolated ionized bubbles, the IGM may contain asymmetric structures sculpted by feedback and local environments \added{(see e.g., \citealp{jalan_probing_2019,zhu_higher_2025} for observations at lower redshifts)}. Radiative transfer simulations that incorporate anisotropic photon escape \citep[e.g.,][]{schwandt_impact_2025} show that narrow ionizing channels can significantly affect ionization morphology, especially at early times. Although such studies do not always predict a directional signal in the 21-cm power spectrum, our results suggest that high-resolution Ly$\alpha$ forest tomography is well suited to reveal these anisotropies, particularly along the most transparent sightlines. On the simulation side, our results also call for a higher resolution in underdense regions to better understand the propagation of ionizing UVB and IGM temperature in the voids.

\section{Summary}\label{sec:summary}

We have presented a detailed analysis of galaxy-IGM correlations near two exceptionally transmissive Ly$\alpha$ forest sightlines at $z \sim 5.7$, based on deep JWST/NIRCam grism observations. Using spectroscopic selection and visual inspection, we identified 28 and 30 [O\,\textsc{iii}] emitters in the J359 and J1306 fields, respectively. Within the redshift intervals corresponding to peak Ly$\alpha$ transmission at $z \simeq 5.7$, we find a significant underdensity of galaxies relative to both field averages and expectations from literature luminosity functions. This confirms earlier reports of galaxy underdensities measured along these lines of sight using LAEs \citep{christenson_relationship_2023}, and supports a scenario in which galaxy-poor regions give rise to the clearest sightlines during the late stages of reionization. At the same time, we note that the transmissive region at $z=5.5$-5.65 toward J1306 appears to coincide with at least an average galaxy density, suggesting that multiple physical pathways may lead to high IGM transmission.


Building on previous comparisons of transmissive sightlines to simulations \citep[e.g.,][]{christenson_relationship_2023,gangolli_correlation_2025}, we carried out a more detailed analysis of our data using forward-modeled observations from FlexRT cosmological simulations. The opacity-density relation indicates that our measurements are consistent with both the scatter predicted by simulations and earlier LAE-based studies. We also examined the transmission-galaxy distance relation, $T(R)$. The 1D $T(R)$ relation shows only modest transmission enhancement at intermediate distances from galaxies, although its constraining power is limited by sample variance. In contrast, the 2D analysis reveals a $\sim 2.5\sigma$ (98th percentile) off-axis enhancement of transmission at large transverse separation and moderate line-of-sight distance from galaxies. Although this feature is not strongly favored by the simulated mean map, it may suggest anisotropic ionized structures aligned with but offset from the QSO sightline. Taken together, the opacity-density relation and $T(R)$ statistics highlight the importance of combining multiple diagnostics to probe reionization geometry \citep[see also, e.g.,][for the power of topology-based analyses]{elbers_persistent_2023}. Future observations that combine wide fields of view with deep spectroscopy will be essential for establishing whether such anisotropic features are common or restricted to rare, low-opacity sightlines.

\begin{acknowledgements}
\added{We than the anonymous reviewer for their insightful and constructive feedback.}
We thank the helpful discussions with Eiichi Egami, Xiangyu Jin, Marcia Rieke, and Fengwu Sun. Y.Z.\ acknowledges support from the NIRCam Science Team contract to the University of Arizona, NAS5-02105. 
G.D.B.\ is supported by JWST Program 4092.
Support for program \#4092 was provided by NASA through a grant from the Space Telescope Science Institute, which is operated by the Association of Universities for Research in Astronomy, Inc., under NASA contract NAS 5-03127.

This work is based on observations made with the NASA/ESA/CSA James Webb Space Telescope. The data were obtained from the Mikulski Archive for Space Telescopes at the Space Telescope Science Institute, which is operated by the Association of Universities for Research in Astronomy, Inc., under NASA contract NAS 5-03127 for JWST. These observations are associated with program \#4092.
\added{Some of the data presented in this paper were obtained from the Mikulski Archive for Space Telescopes (MAST) at the Space Telescope Science Institute. The specific observations analyzed can be accessed via \dataset[doi: 10.17909/4z7r-ef97]{\doi{10.17909/4z7r-ef97}}. STScI is operated by the Association of Universities for Research in Astronomy, Inc., under NASA contract NAS5-26555. Support to MAST for these data is provided by the NASA Office of Space Science via grant NAG5–7584 and by other grants and contracts.}


This manuscript benefited from grammar checking and proofreading using ChatGPT \citep{openai_chatgpt_2024}. 

\end{acknowledgements}


\vspace{5mm}
\facilities{JWST; MAST}

\software{
{\tt astropy} \citep{astropy_collaboration_astropy_2013,astropy_collaboration_astropy_2018,astropy_collaboration_astropy_2022},
{\tt JWST Calibration Pipeline} \citep{bushouse_jwst_2022},
{\tt scipy} \citep{virtanen_scipy_2020}
}

\appendix

\section{Completeness and Contamination} \label{sec:completeness}

To quantify the completeness of our [O\,\textsc{iii}] emitter selection, we perform an injection-recovery test using $10^7$ mock [O\,\textsc{iii}] sources. These mock emitters are randomly assigned redshifts from the list $z = 5.1$ to $6.2$ in steps of $\Delta z = 0.1$, fluxes drawn from a log-spaced range between $10^{-18}$ and $10^{-15}$~erg~s$^{-1}$~cm$^{-2}$, and line widths sampled logarithmically from 100 to $1000$~km~s$^{-1}$. Each mock emitter is assigned a random 2D position on the full mosaic, and the local noise properties are extracted from the real noise maps at that location. These synthetic spectra are then passed through the same automated [O\,\textsc{iii}] identification pipeline used for the real data. 
Because the injected spectra contain genuine [O\,\textsc{iii}] doublets, they would almost always ($\gtrsim 95\%$) be accepted during visual inspection, which in practice mainly rejects noise artifacts in the real data. We therefore do not apply visual inspection to the mock sample.

\begin{figure}[!ht]
    \centering
    \includegraphics[width=1.0\linewidth]{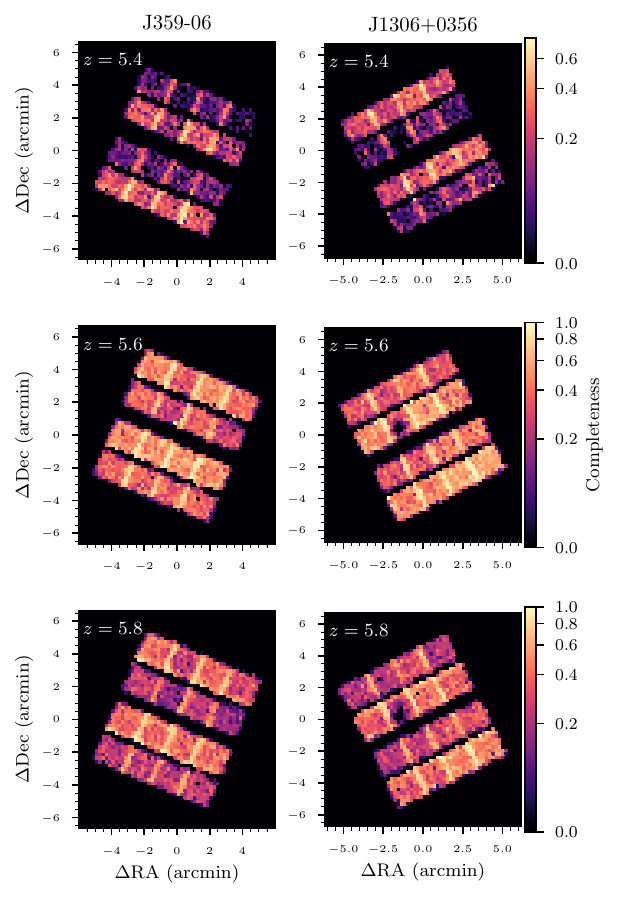}
    \caption{
Two-dimensional completeness maps for detecting [O\,\textsc{iii}]~5008\,\AA\ emitters above a luminosity threshold of $\log_{10}(L_{5008}/[\mathrm{erg\,s}^{-1}]) > 42.1$ across the J359$-$06 (left column) and J1306$+$0356 (right column) NIRCam fields. Each row shows the completeness in a different redshift slice centered at $z = 5.4$, $z = 5.6$, and $z = 5.8$, respectively. 
}
    \label{fig:comp2d}
\end{figure}

Figure~\ref{fig:comp2d} shows the two-dimensional completeness maps in the J359 and J1306 fields for galaxies with $\log L_{5008} > 42.1$. Regions of low completeness correspond to areas where [O\,\textsc{iii}]~$\lambda5008$ falls outside the grism wavelength coverage, a limitation imposed by the reduced overlap between NIRCam pointings. This observing strategy was chosen to maximize transverse distance coverage from the field center. As tested in Appendix~\ref{sec:convergence}, applying the exact survey footprint (including the central gaps around the QSO positions) has negligible impact on the shape of both the 1D and 2D galaxy--IGM cross-correlations. Enhanced completeness in the fields arises from overlap between pointings, which leads to lower noise and improved recovery. Small holes in the 2D maps mark the positions of bright stars, where line injection was not performed.

\begin{figure}[!ht]
    \centering
    \includegraphics[width=1.0\linewidth]{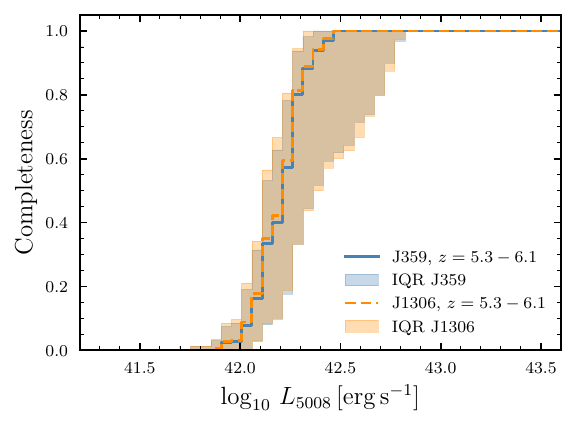}
    \caption{
One-dimensional completeness curves for [O\,\textsc{iii}]~5008\,\AA\ emitter detection as a function of logarithmic luminosity, $\log_{10}(L_{5008}/[\mathrm{erg\,s}^{-1}])$, for the J359 (solid blue) and J1306 (dashed orange) NIRCam fields. The curves represent the median completeness within the redshift range $5.3 < z < 6.1$, while the shaded regions show the interquartile range (IQR) across all spatial positions and redshift slices. The 80\% completeness thresholds are nearly identical between the fields at $\log_{10}(L_{5008}) \approx 42.3$.
}
    \label{fig:completeness_curve}
\end{figure}

Figure~\ref{fig:completeness_curve} shows the 1D completeness curves as a function of [O\,\textsc{iii}]~$\lambda5008$ luminosity, averaged over each field. The J359 and J1306 fields show nearly identical completeness behavior due to similar configurations and exposure depths. The completeness rises steeply with luminosity, reaching approximately 40\% at $\log L_{5008} = 42.1$ and 80\% at $\log L_{5008} = 42.3$.

To estimate the contamination rate, we also applied the same selection pipeline to search for artificial line pairs with incorrect wavelength separation (e.g., $\lambda = 4900$ and $5008$~\AA{} instead of the physical $\lambda = 4960$ and $5008$~\AA{}). We then applied visual inspection to these false-positive candidates. We find that the contamination rate after visual inspection is low, typically $<10\%$ across both fields.

\section{Metal Absorption Lines Corresponding to the Overdensity} \label{sec:metal}

\begin{figure}[!ht]
    \centering
    \vspace{1em}
    \includegraphics[width=1.0\linewidth]{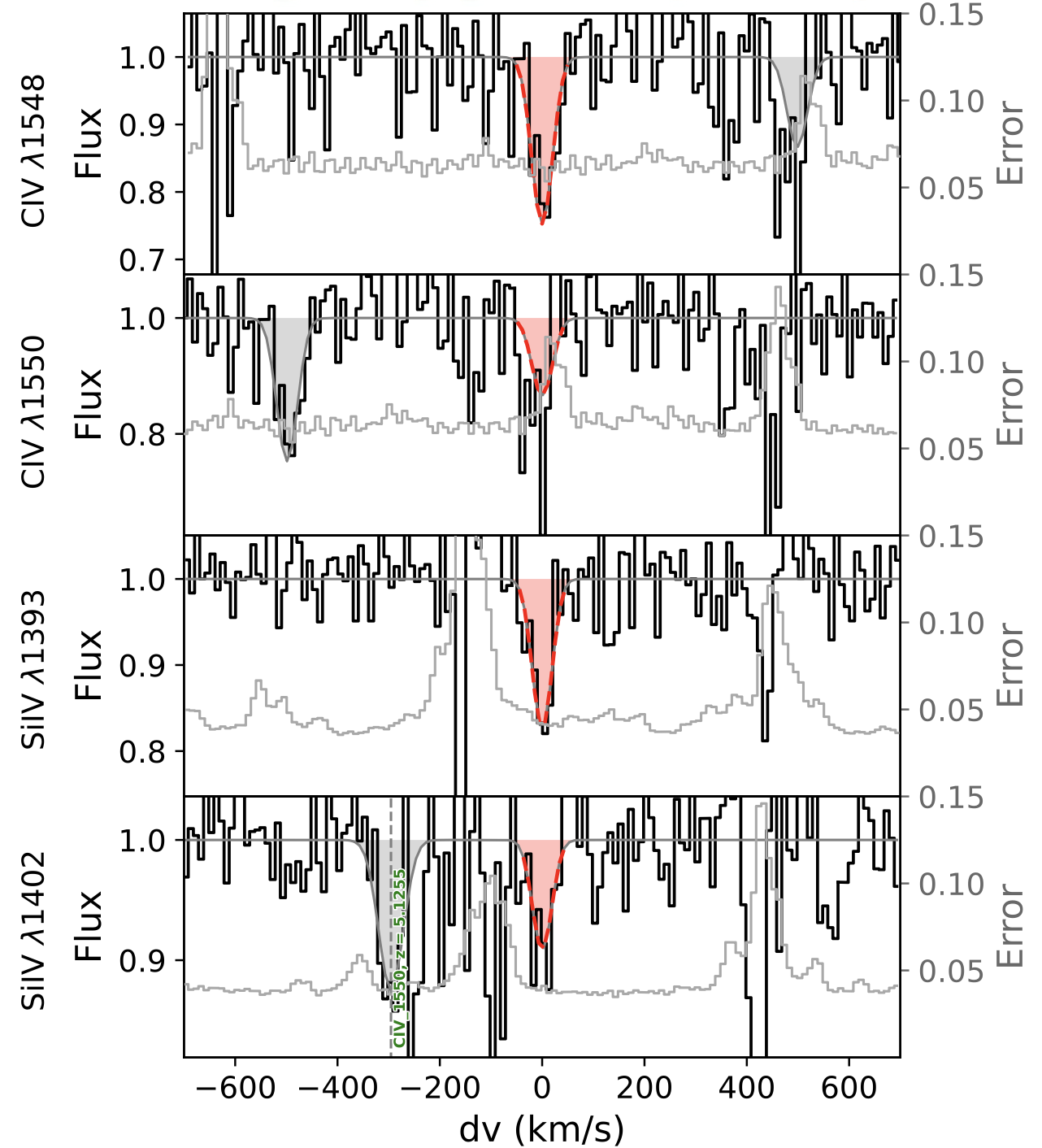}
    \caption{High-ionization metal absorption lines detected at $z = 5.779$ in the spectrum of PSOJ359--06. The panels show C\,\textsc{iv} $\lambda\lambda$1548, 1550 and Si\,\textsc{iv} $\lambda\lambda$1393, 1402, with absorption profiles (black) and associated errors (gray). Shaded red regions show Voigt profile fits from \citet{davies_xqr-30_2023}. This figure is adapted from the publicly released XQR-30 metal line catalog at \url{https://github.com/XQR-30/Metal-catalogue}.}
    \label{fig:metal}
\end{figure}

Figure~\ref{fig:metal} shows the detection of high-ionization metal absorption lines: C\,\textsc{iv} and Si\,\textsc{iv} at $z = 5.779$ along the sightline toward PSOJ359--06, coinciding with the redshift of the [O\,\textsc{iii}] overdensity discussed in Section~\ref{sec:results}. These lines are identified in the high-resolution XQR-30 spectrum \citep{davies_xqr-30_2023, davies_examining_2023}.

The absorption system exhibits significant detections of both C\,\textsc{iv} and Si\,\textsc{iv}, with rest-frame equivalent widths of $W_{1548} = 0.06$~\AA{} and $W_{1393} = 0.04$~\AA{}, respectively. Voigt profile fitting yields column densities of $\log N_{\mathrm{C\,IV}} = 13.26 \pm 0.08 \,\rm cm^{-2}$ and $\log N_{\mathrm{Si\,IV}} = 12.68 \pm 0.08 \,\rm cm^{-2}$. These measurements indicate the presence of metal-enriched and highly ionized gas associated with the galaxy overdensity at $z = 5.78$. However, despite the high ionization state, this region shows no detectable Ly$\alpha$ transmission in the forest, suggesting the presence of residual high column density H\,\textsc{i}, typical of metal absorbers.

\section{Convergence Tests} \label{sec:convergence}

To assess the robustness of the 1D and 2D IGM-galaxy cross-correlation functions (CCFs) presented in Figures~\ref{fig:1d_opacity_density} and \ref{fig:2d_obs}, we perform a suite of convergence tests using mock observations generated from the \texttt{FlexRT} simulation. These tests examine the effects of varying redshift, field of view (FOV), and $M_{\mathrm{UV}}$ selection thresholds for halos used in the stacking analysis.

For each test, we generate 1D and 2D stacked transmission maps across a grid of redshifts ($z = 5.2$-$6.0$), FOVs (matching the observed geometry and an extended version with $\gtrsim 1.5\,\lambda_{\mathrm{mfp}}$ in radius), and galaxy number densities corresponding to different $M_{\mathrm{UV}}$ limits (bright: $\sim$15 halos per field, medium: $\sim$30, faint: $\sim$60). 
To mimic the incompleteness near the QSO positions, we apply an inner mask following the true NIRCam footprint, which includes empty stripes in the mosaic.

\begin{figure*}[!ht]
    \centering 
    \includegraphics[width=1\linewidth]{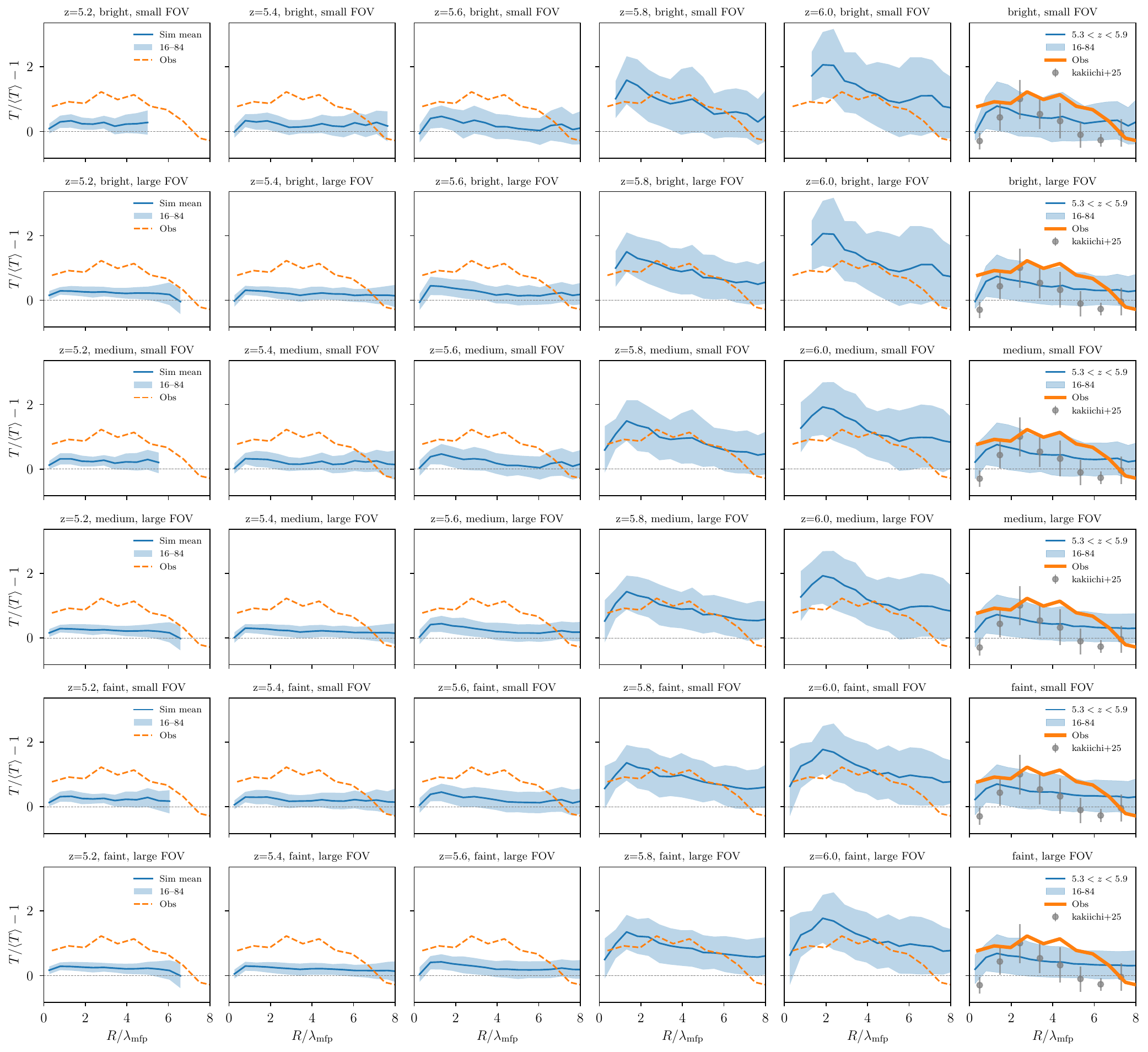}
    \caption{
One-dimensional IGM-galaxy cross-correlation functions from mock sightlines in the FlexRT simulation, shown for different galaxy selection thresholds and field-of-view (FOV) sizes. Each row corresponds to a different UV magnitude cut: bright ($M_{\mathrm{UV}}$ limit yielding $\sim15$ galaxies per field), medium ($\sim30$ galaxies, matching our observed sample size), and faint ($\sim60$ galaxies). Each column from left to right varies in redshift and survey geometry: from $z=5.2$ to $z=6.0$, and for small (matching the observed FOV) or large ($>1.5 \lambda_{\mathrm{mfp}}$) survey volumes. The final column stacks the results over $5.3<z<5.9$ to directly compare with the observed 1D cross-correlation (orange). Shaded regions indicate the 16th-84th percentile range from 10000 mock realizations.}
\label{fig:app_1d}
\end{figure*}

Figure~\ref{fig:app_1d} presents the 1D cross-correlation results. We find that the qualitative trends are consistent across different FOV sizes and galaxy sample depths. The amplitude of the excess transmission, $T/\langle T \rangle - 1$, becomes more prominent with increasing redshift, suggesting stronger fluctuations in the ionizing background. This is consistent with a more homogeneous UVB at lower redshifts, where Ly$\alpha$ transmission is increasingly governed by the large-scale density field rather than by local sources \citep[see e.g.,][]{kashino_eiger_2025}. 

\begin{figure*}[!ht]
    \centering 
    \includegraphics[width=1\linewidth]{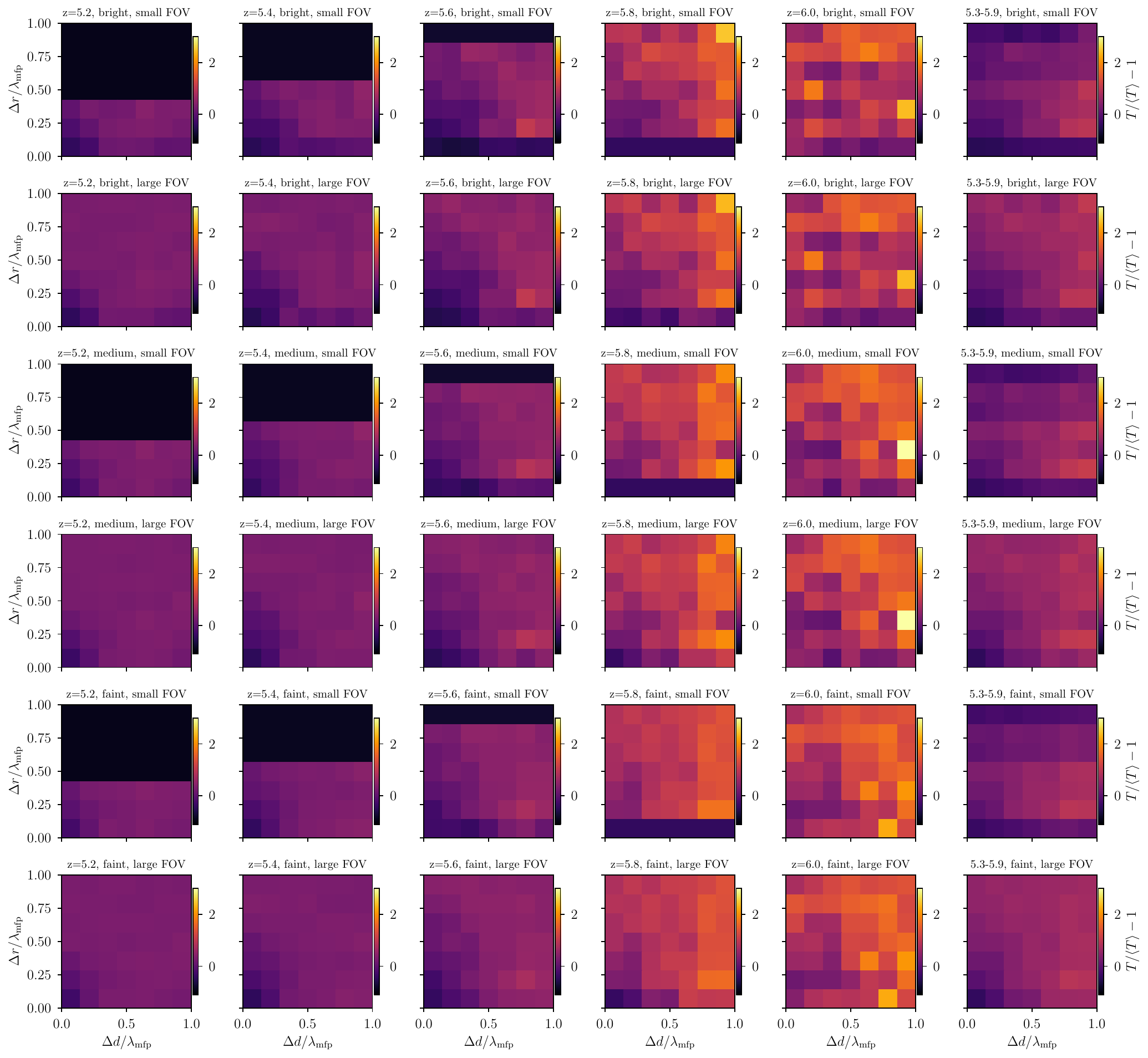}
    \caption{
Two-dimensional IGM-galaxy cross-correlation maps from the FlexRT simulation, analogous to Figure~\ref{fig:app_1d} but shown in 2D. Each panel displays the normalized transmission excess $T/\langle T \rangle - 1$ as a function of line-of-sight ($\Delta d$) and transverse ($\Delta r$) separation from galaxies, both normalized by the mean free path ($\lambda_{\mathrm{mfp}}$). Rows indicate different $M_{\mathrm{UV}}$ thresholds and columns span redshifts from $z=5.2$ to $z=6.0$, with the final column stacking the range $5.3<z<5.9$. 
}
    \label{fig:app_2d}
\end{figure*}

Figure~\ref{fig:app_2d} shows the corresponding 2D CCF maps. These exhibit the same trends seen in the 1D case: higher redshifts yield stronger and more extended fluctuations in $T/\langle T \rangle - 1$. Changes in the FOV and $M_{\mathrm{UV}}$ threshold do not significantly alter the overall morphology of the stacked transmission maps. Importantly, none of the mock realizations reproduce the prominent horizontal band of enhanced transmission at $\Delta r \sim 0.8\,\lambda_{\mathrm{mfp}}$ seen in the observed 2D map (Figure~\ref{fig:2d_obs}).

\bibliographystyle{aasjournalv7}

\end{document}